\documentclass[twocolumn,prb]{revtex4}

\usepackage{graphicx}

\begin{document}


\title{Mechanism and Scalability in Resistive Switching of 
Metal-Pr$_{0.7}$Ca$_{0.3}$MnO$_{3}$ Interface}

\author{S.~Tsui}
\author{Y.~Q.~Wang}
\author{Y.~Y.~Xue}
\affiliation{Department of Physics and Texas Center for Superconductivity, 
University of Houston, 202 Houston Science Center, Houston, Texas 77204-5002}
\author{C.~W.~Chu}
\affiliation{Department of Physics and Texas Center for Superconductivity, 
University of Houston, 202 Houston Science Center, Houston, Texas 77204-5002; 
Lawrence Berkeley National Laboratory, 1 Cyclotron Road, 
Berkeley, California 94720; and Hong Kong University of Science and Technology, 
Hong Kong}

\date{\today}

\begin{abstract}
The polarity-dependent resistive-switching across metal-Pr$_{0.7}$Ca$_{0.3}$MnO$_{3}$ 
interfaces is investigated.  The data suggest that shallow defects in the interface 
dominate the switching. Their density and fluctuation, therefore, will ultimately limit 
the device size. While the defects generated/annihilated by the pulses and the associated 
carrier depletion seem to play the major role at lower defect density, the defect 
correlations and their associated hopping ranges appear to dominate at higher defect 
density. Therefore, the switching characteristics, especially the size-scalability, may be 
altered through interface treatments.
\end{abstract}

\maketitle

The renewed interest in various resistive switching phenomena is largely driven by recent 
market demands for nano-sized nonvolatile memory devices.\cite{wei} While the current boom 
of consumer electronics may largely be attributed to the successful miniaturization of 
both FLASH chips and mini hard drives, cheaper and smaller devices are called for. Various 
resistive hysteretic phenomena are consequently studied with the hope that the size 
limitations associated with the related physics/chemistry/technology might be less 
severe.\cite{was} Our limited knowledge about the mechanisms so far, however, makes the 
evaluation difficult. This is especially true for the switching across 
metal-Pr$_{0.7}$Ca$_{0.3}$MnO$_{3}$ (PCMO) interfaces.\cite{liu,bai,saw,oda} Several models, 
\textit{i.e.} bulk phase-separation,\cite{liu} carrier-trapping in pre-existing metallic 
domains,\cite{roz,roz2} and field-induced lattice defects,\cite{bai} have been 
proposed. Each possesses its own distinguishable size-limitation, \textit{e.g.} the 
statistics of the associated local mesostructures. Here, we report our mechanism 
investigation through both the trapped-carrier distribution and their hopping range. 
Our data suggest that the characteristics may largely be engineered through the 
mesostructure of the interfacial defects.

Bulk PCMO, in great contrast with well known semiconductors, has a rather high nominal 
carrier concentration with its high resistivity mainly attributed to hopping 
barriers.\cite{hwa} Local defects, therefore, appear as a natural cause of the resistive 
switching. Following this line of reasoning, a domain model has recently attracted much 
attention.\cite{roz,roz2} There, a tunneling from the electrode to some pre-existing 
interfacial metallic domains has been assumed to be the dominant process. Consequently, 
the carrier-occupation in the domains may change with the carrier-trapping during the 
write pulses, and cause the $R$-switch between an on (low resistance) and an off 
(high resistance) state. This is realized through either the change of the tunneling 
probability\cite{roz} or a doping-induced metal-insulator transition.\cite{roz2} Useful 
devices based on this mechanism, therefore, should typically be much larger than these 
interfacial domains. It is interesting to note that even if the ``domains'' can be reduced to 
individual lattice defects (or small clusters) as in the proposed defect modification 
model,\cite{bai} the fluctuation (inhomogeneity) of the defect density still sets a limit 
for the size scalability just like the dopant fluctuation in Si nano-devices.\cite{mei} 
The defect (domain) density, therefore, requires exploration, and the interfacial 
capacitance, $C(\omega)$, can serve to distinguish between these models.

While the domain model may simulate very divergent dc $I-V$ characteristics by adjusting 
the fitting parameters,\cite{roz} the measured capacitance is expected to be 
$\omega$-independent with $C_{on} = C_{off}$ ($C_{on} > C_{off}$) for tunneling-probability 
(metal-insulator transition) scenarios, where $C_{on}$ and $C_{off}$ are $C(\omega)$ in the 
on and off states, respectively. In the defect modification model, however, the $C(\omega)$ 
measures the net trapped carriers in the interface, \textit{i.e.} with 
$R_{off}/R_{on} \approx C_{off}/C_{on}$ in the space-charge-limited-current (SCLC) region, 
where $R_{on}$ and $R_{off}$ are $R(\omega)$ in the on and off states, respectively. It should 
be noted that the carriers trapped behind a hopping barrier $V_{hop}$ respond to a 
step-disturbance as $\exp[-v_{0}t \cdot \exp(-V_{hop}/k_{B}T)]$. Therefore, 
$C(\omega) \propto \int\limits_{\omega}^{\infty}k_{B}T[dN/dV_{hop}]/\omega d \omega$, \textit{i.e.} 
the density of states (DOS) at the Fermi level for the defects with the hopping barrier 
lower than $V_{hop} = k_{B}T \ln (v_{0}/\omega)$,  where $k_{B}$ and $v_{0} \approx 
10^{12}$ sec$^{-1}$ are the Boltzmann constant and the trial frequency, respectively. An 
experimental challenge, however, exists in separating the interfacial $C(\omega)$ from the 
bulk contribution. We have previously reported a preliminary result for an Ag-PCMO 
interface\cite{bai} through the traditional Cole-Cole procedure, which assumes that all $C$'s 
and $R$'s are $\omega$-independent, under a standard two-leads measurement configuration. 
In the SCLC region, the observed $C_{off} = 2.2$ nF and $C_{on} = 1.6$ nF clearly contradict 
the domain model but qualitatively agree with the defect modification model. The data, 
however, also raise a serious concern about the size-limitation. In a sense, the observed 
$C/q \approx 10^{12}$ electrons$/$cm$^2$, where $q =1.6 \cdot 10^{-19} C$ is the electron 
charge, is a measure of the trapped carriers (or the shallow defects near the Fermi level), 
and puts a statistical size-limit on the order of 10--100 nm if the switching is due to a 
change in defect density. Therefore, a possible route to further minimizing the device size 
would be to increase the $C(\omega)$. A dilemma, however, arises: denser shallow defects 
would also enhance the thermal excitation. Across a $C(\omega)$ threshold, the SCLC may not 
be reachable, and the denser defects at the off states, functioning as donors, might even 
enhance the conductivity. It is therefore even unclear whether samples with a much larger 
$C(\omega)$ are switchable. 

\begin{figure}
\includegraphics{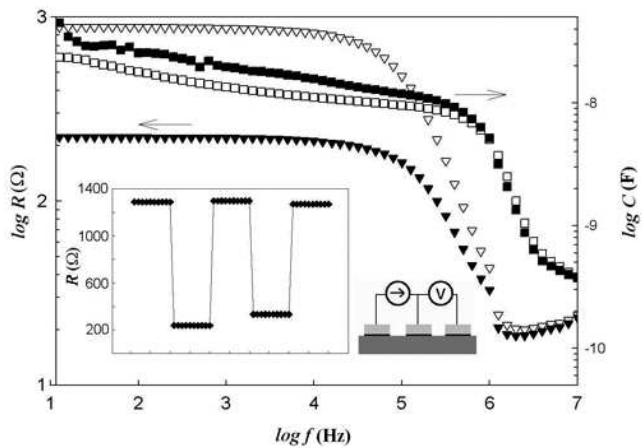}
\caption{\label{fig:fig1}$R(\omega)$ (triangle symbols) and $C(\omega)$ (square symbols) of a 
Au-PCMO setup at on- (solid) and off- (open) states. Bottom left inset: the switching series. 
Bottom right inset: three-lead setup.}
\end{figure}

We finally found several metal electrode-PCMO film configurations, with the low-$\omega$ 
interfacial $C > 100$ nF$/$mm$^2$, a value one order of magnitude higher, although the 
conditions for reproducibly synthesizing such samples are not yet clear. Repeatable 
switching has been obtained (bottom left inset, Fig.~\ref{fig:fig1}).  The Cole-Cole plot, 
however, shows 
that the $C_{off}/C_{on}$ is even smaller than 1, while $R_{off}/R_{on} > 2$, a scenario 
closer to the domain model. To explore the issue, a new procedure is developed to directly 
measure the complex interfacial admittance $1/R(\omega)-i\omega C(\omega)$. This is done by 
extending the previous three-leads $R$-measurement to the off-phase part through a 
Solartron SI 1260 impedance/gain-phase analyzer (bottom right inset, Fig.~\ref{fig:fig1}). 
Resistors/capacitors networks were used to verify that the phase uncertainty is less than 
$0.1^0$. Both $R(\omega)$ and $C(\omega)$ of the interface, therefore, can be accurately 
deduced over $10^2$--$10^7$ Hz. 

\begin{figure}
\includegraphics{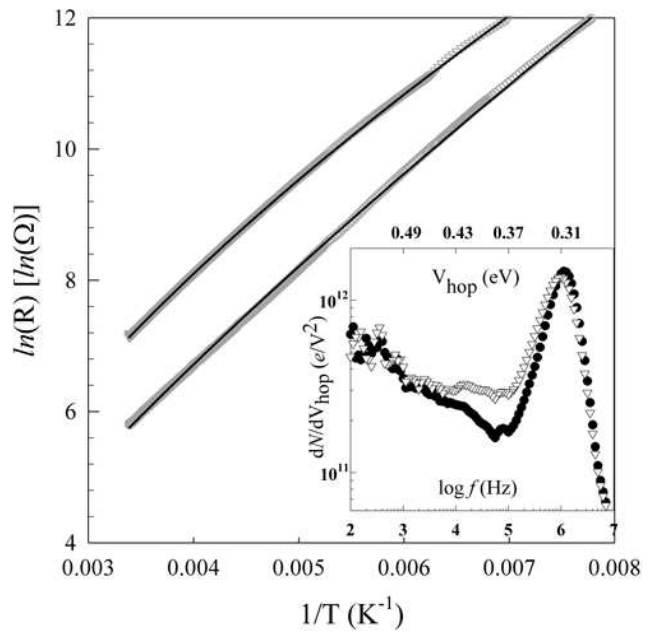}
\caption{\label{fig:fig2}$R(T)$ between 150 and 300 K. Top (bottom) grey symbols: 
data at off- (on-) states. Lines: VRH fits. Inset: the calculated defect distribution in 
the 0.1 mm$^2$ interface layer. Open triangles: the on state. Solid circles: the off state.}
\end{figure}

The $C(\omega)$ and the $R(\omega)$ observed undergo a step-like jump over a narrow range 
of 0.1 and 3 MHz, with the $C$-jump occurring at higher $\omega$ (Fig.~\ref{fig:fig1}). 
This is similar 
to the Maxwell-Wagner relaxation,\cite{yu} but differs from dielectric Debye 
relaxation.\cite{jon} Carrier polarization, \textit{i.e.} trapping and hopping in 
disordered solids,\cite{dyr} appears to be a natural interpretation. In such models, only 
the hopping paths with all barriers $V_{hop} \ll k_{B}T \ln (\omega/v_{0})$ contribute to 
the apparent conductivity $1/R(\omega)$, and the $\omega dC(\omega)/d\omega$ measures the 
defect distribution against the hopping barriers, $dN/dV_{hop}$, at the Fermi level. The 
jumps (Fig.~\ref{fig:fig1}), therefore, suggest a defect mesostructure with many short 
conducting 
clusters (domains) with the intra-cluster $V_{hop} < 0.3$ eV separated by slightly higher 
inter-cluster barriers, \textit{i.e.} around 0.4 eV (inset, Fig.~\ref{fig:fig2}). It is 
interesting to 
note that the $R_{on}$ and $R_{off}$ are distinct, \textit{i.e.} switchable, below 
0.1--1 MHz, but the corresponding $dN/dV_{hop}$ is practically the same as indicated by 
the Cole-Cole plot. This is very different from the samples with smaller $C(\omega)$, 
although both show higher $R$ after positive pulses.   Also, the same $R(\omega)$ above 
1 MHz at both on and off states indicates a limitation on the read speed in future potential 
applications.

It should be pointed out that such strong dispersions of $R(\omega)$ and $C(\omega)$ suggest 
that the domain model is also in disagreement with the data. The interfacial hopping in the 
model has been attributed to a \textit{single} barrier layer between the electrode and the 
proposed interfacial domain, and should therefore be $\omega$-independent for 
$\omega \ll v_0$. Thus, a more complicated defect network should be invoked to accommodate 
the dispersions. It is also interesting to note that the thermo-produced carriers can be 
directly deduced from the $C(\omega)$ observed. The calculated value, $> 10^{17}$ sec$^{-1}$ 
below 100 kHz within the interfacial layer, is far higher than the reported injected current, 
$\le 5 \cdot 10^{14}$ sec$^{-1}$, at the SCLC region.\cite{bai} This supports the above 
discussion for samples with large $C(\omega)$, and suggests that the switching mechanism can 
be modified through interface engineering. 

To answer the key question of why the switching can still occur, $R(T)$ was measured at 
both on- and off-states (Fig.~\ref{fig:fig2}). While the $R(T)$'s appear to be almost parallel, which 
makes the switching difficult to understand if $C_{off} \approx C_{on}$, a closer 
examination shows that $R = a \cdot T \cdot \exp[-(T_{0}/T)^{\gamma}]$ in the 
variable-range-hopping (VRH) formulation might be a better description, as suggested by the 
slight curvatures in Fig.~\ref{fig:fig2}. For this particular sample, the fitting 
parameters are 
$\gamma = 0.85$ (0.61) and $T_{0} = 2089$ K (8353 K) for the on (off) states, respectively. 
The $\gamma$ and $T_{0}$ are traditionally associated with $1/(1+d)$ and $1/$DOS, where 
$d$ is the dimensionality.\cite{san} 
The changes in both $\gamma$ and $T_{0}$, therefore, tentatively suggest that the arrangement 
of the defect-structures are changed.  The switching, in such a case, might be more about 
the local structure than the average defect density.

\begin{figure}
\includegraphics{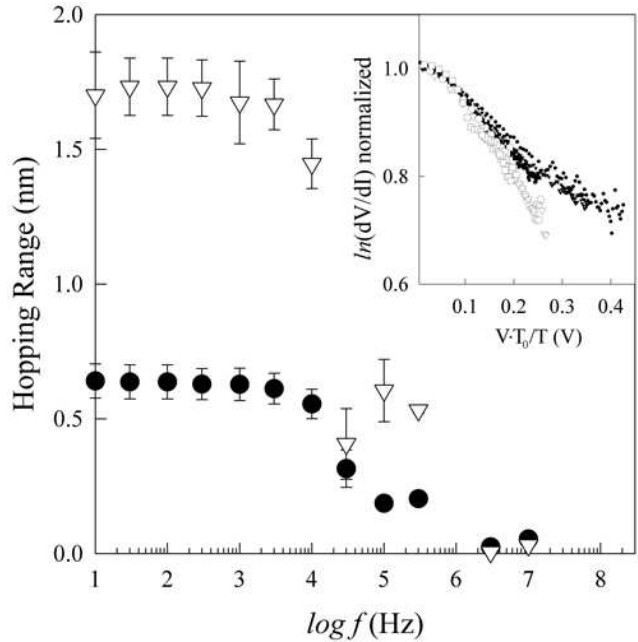}
\caption{\label{fig:fig3}Hopping range with respect to frequency for both off 
(open triangles) and on 
(solid circles) states. The switch is determined by low frequency inter-cluster hopping, 
whereas at high frequencies, percolative intra-cluster hopping occurs.  
Inset: raw $I-V$ data at several temperatures show the scaling of Mott's formula, where 
$T_{0} = 300$ K.}
\end{figure}

To verify this assumption, the small-signal ac $R(\omega)$ at various dc biases and 
temperatures was measured (inset, Fig.~\ref{fig:fig3}). The normalized $I-V$ 
characteristics between 8 and 
55 $^{\circ}$C are scaled into a single trace as suggested by Mott's formula of 
$R/R(E=0) \propto \exp(-l \cdot E/k_{B}T)^{\gamma}$ or $\exp(-l \cdot E/k_{B}T)$,\cite{mot} 
where $l$ and $E$ are the hopping range and the electric field, respectively. The physical 
picture, \textit{i.e.} the longer the hopping range, the larger the field effects will be, is 
straightforward and independent of the hopping details, although the absolute value may be 
affected by the 10 nm assumed thickness\cite{bai} of the interfacial layer. Our data, 
therefore, demonstrate that the hopping range is longer at the off state for $\omega < 1$ 
MHz. It is also interesting to note the unusually small $l$ for $\omega > 1$ MHz 
(Fig.~\ref{fig:fig1}, \ref{fig:fig3}). This may appear only if the associated hopping 
barriers are negligible, which 
is further suggestive of local defect mesostructure. With high enough defect density, the 
formation of local defect-correlations becomes a means to modify the switching performance. 
It should be noted that, in order to accommodate the observed $C_{off}/C_{on}$, the 
conductance may be dominated by only a few percolation paths, such that simply 
enhancing $C(\omega)$ may not necessarily reduce the size limitations. Direct observations 
of such structures are called for and planned. 

In conclusion, we have shown that the interfacial resistive switch found in metal-PCMO 
systems is a defect-mediated process. Whereas a defect density alteration occurs in samples 
with $C(\omega < 10^{5}$ Hz) $\le 100$ nF/mm$^{2}$, we propose that the change in defect 
mesostructure may cause switching in the samples with larger $C(\omega)$. Further 
investigation of the parameters controlling the defect distribution will shed light on how 
best to proceed with the nanoscaling and benchmarking of future device applications.

\begin{acknowledgments}
The authors thank Prof. J. Miller and Dr. N. Nawarathna for the use of and technical 
assistance with the impedance analyzer. The work in Houston is supported in part by 
NSF Grant No. DMR-9804325, the T.~L.~L. Temple Foundation, the John J. and Rebecca 
Moores Endowment, the Robert A. Welch Foundation, and the State of Texas 
through the Texas Center for 
Superconductivity at the University of Houston; and at Lawrence Berkeley 
Laboratory by the Director, Office of Science, Office of Basic Energy Sciences, 
Division of Materials Sciences and Engineering of the U.S. Department of Energy 
under Contract No. DE-AC03-76SF00098.
\end{acknowledgments}


\end{document}